\documentclass[aps,prl,twocolumn,showpacs,psfig,superscriptaddress,longbibliography]{revtex4-1}

\usepackage{textcomp}
\usepackage{times}
\usepackage{graphicx}
\usepackage{float}
\usepackage{latexsym,amsmath,amssymb,bm,euscript}
\usepackage{color}
\usepackage{subfigure}
\usepackage{epstopdf}
\usepackage[colorlinks=true,linkcolor=black,citecolor=black,urlcolor=black]{hyperref}
\usepackage{hyperref}
\usepackage{soul}
\usepackage[normalem]{ulem}
\usepackage{mathrsfs}
\usepackage{amsmath}
\usepackage{lettrine}
\usepackage{xspace}
\usepackage{textcomp}

\def\para{\ensuremath{/\kern -0.8em /}\xspace}

\begin{document}

\title{The observation of quantum fluctuations in a kagome Heisenberg antiferromagnet}

\author{Fangjun Lu}
\thanks{These authors contributed equally}
\affiliation{Wuhan National High Magnetic Field Center and School of Physics, Huazhong University of Science and Technology, 430074 Wuhan, China}

\author{Long Yuan}
\thanks{These authors contributed equally}
\affiliation{Wuhan National High Magnetic Field Center and School of Physics, Huazhong University of Science and Technology, 430074 Wuhan, China}

\author{Jian Zhang}
\affiliation{Wuhan National High Magnetic Field Center and School of Physics, Huazhong University of Science and Technology, 430074 Wuhan, China}

\author{Boqiang Li}
\affiliation{Wuhan National High Magnetic Field Center and School of Physics, Huazhong University of Science and Technology, 430074 Wuhan, China}

\author{Yongkang Luo}
\email{mpzslyk@hust.edu.cn}
\affiliation{Wuhan National High Magnetic Field Center and School of Physics, Huazhong University of Science and Technology, 430074 Wuhan, China}

\author{Yuesheng Li}
\email{yuesheng\_li@hust.edu.cn}
\affiliation{Wuhan National High Magnetic Field Center and School of Physics, Huazhong University of Science and Technology, 430074 Wuhan, China}

\begin{abstract}
\end{abstract}

\date{\today}
\maketitle

\noindent \textbf{Abstract}

\textbf{The search for the experimental evidence of quantum spin liquid (QSL) states is critical but extremely challenging, as the quenched interaction randomness introduced by structural imperfection is usually inevitable in real materials. YCu$_3$(OH)$_{6.5}$Br$_{2.5}$ (YCOB) is a spin-1/2 kagome Heisenberg antiferromagnet (KHA) with strong coupling of $\langle J_1\rangle\sim$ 51 K but without conventional magnetic freezing down to 50 mK $\sim$ 0.001$\langle J_1\rangle$. Here, we report a Br nuclear magnetic resonance (NMR) study of the local spin susceptibility and dynamics on the single crystal of YCOB. The temperature dependence of NMR main-line shifts and broadening can be well understood within the frame of the KHA model with randomly distributed hexagons of alternate exchanges, compatible with the formation of a randomness-induced QSL state at low temperatures. The in-plane spin fluctuations as measured by the spin-lattice relaxation rates ($1/T_1$) exhibit a weak temperature dependence down to $T$ $\sim$ 0.03$\langle J_1\rangle$. Our results demonstrate that the majority of spins remain highly fluctuating at low temperatures despite the quenched disorder in YCOB.
}

\bigskip \noindent {\bf Introduction}\\
\lettrine[lines=2, findent=3pt, nindent=0pt]{Q}
{\textnormal{uantum}} spin liquid (QSL) is a state of matter that exhibits exotic fractional excitations and long-range entanglement without symmetry breaking~\cite{balents2010spin,Broholmeaay0668,kasahara2018majorana,RevModPhys.89.025003}. Since Anderson’s proposal of the prototype, i.e., resonating-valence-bond (RVB) state, in 1973~\cite{anderson1973resonating}, QSL has been attracting researchers for decades,  due to its key role in understanding high-temperature superconductivity~\cite{anderson1987resonating} and the possible realization of the topological quantum computation~\cite{nayak2008non}. Experimentally, many prominent two-dimensional QSL candidate compounds have been extensively studied (the one-dimensional scenario of QSL is qualitatively different~\cite{Broholmeaay0668}), including the kagome-lattice ZnCu$_3$(OH)$_6$Cl$_2$ (herbertsmithite)~\cite{Shores2010A,PhysRevLett.98.107204,PhysRevB.76.132411,PhysRevLett.100.157205,han2012fractionalized,PhysRevB.94.060409,PhysRevLett.100.087202,Fu2015Evidence,Khuntia2020Gapless,RevModPhys.88.041002}, triangular-lattice $\kappa$-(ET)$_2$Cu$_2$(CN)$_3$~\cite{PhysRevLett.91.107001,Yamashita2008Thermodynamic}, EtMe$_3$Sb[Pd(dmit)$_2$]$_2$~\cite{PhysRevB.77.104413,Itou2010Instability}, YbMgGaO$_4$~\cite{li2015gapless,PhysRevLett.115.167203}, etc., all of which generally exhibit gapless QSL behaviors~\cite{PhysRevLett.98.107204,PhysRevLett.100.157205,PhysRevLett.100.087202,Khuntia2020Gapless,PhysRevLett.91.107001,Yamashita2008Thermodynamic,PhysRevB.77.104413,li2015gapless,PhysRevLett.117.097201,shen2016spinon,paddison2016continuous,PhysRevLett.122.137201}, but without evident magnetic thermal conductivity~\cite{huang2021heat,murayama2021universal,PhysRevLett.123.247204,PhysRevX.9.041051,PhysRevLett.117.267202}.

Despite the progress, the existing experimental evidence for QSL remains circumstantial and strongly depends on theoretical interpretation. The root cause lies in the quenched interaction randomness introduced by structural imperfection that is inevitable in all real materials~\cite{Broholmeaay0668,RevModPhys.88.041002,PhysRevB.92.134407}. Therefore, great efforts are being devoted to exploring for ultrahigh-quality candidate materials, which is extremely challenging~\cite{Broholmeaay0668}. On the other hand, disorder-free QSL, even if successfully prepared, is usually very fragile. For instance, the most frustrated kagome Heisenberg antiferromagnet (KHA) falls back to conventional long-range magnetic ordering in the presence of a weak next-nearest-neighbor coupling $|J_2|$ $\geq$ $0.03J_1$~\cite{PhysRevLett.118.137202} or Dzyaloshinsky-Moriya interaction $|D|$ $\geq$ $0.012J_1$ ~\cite{PhysRevB.98.224414}. These constrictions further compress the ``living space" of disorder-free perfect QSL compounds.

Alternatively but more realistically, one could first find out whether the inherent randomness is fatal or vital to the QSL physics~\cite{RevModPhys.89.025003}. In fact, this same question can also be raised for high-temperature superconductivity, as it is generally believed that Cooper pairs naturally form once the RVB states are charged upon chemical doping~\cite{anderson1987resonating,RevModPhys.78.17}. The presence of quenched vacancies in the KHA can lead to a valence bond glass ground state (GS)~\cite{PhysRevLett.104.177203}. Further, Kawamura et al. found that randomness-induced QSL GSs instead of spin glasses form in both KHA and triangular Heisenberg antiferromagnet with strong bond randomness, $\Delta J/J_1$ $\geq$ 0.4 and 0.6~\cite{2014Quantum,2014Quantumspin}, respectively, which may explain the gapless behaviors observed in ZnCu$_3$(OH)$_6$Cl$_2$ , $\kappa$-(ET)$_2$Cu$_2$(CN)$_3$, EtMe$_3$Sb[Pd(dmit)$_2$]$_2$~\cite{PhysRevB.92.134407}, etc. Later similar scenarios have been generally applied to the gapless QSL behaviors observed in the strongly-spin-orbital-coupled triangular-lattice YbMgGaO$_4$~\cite{PhysRevLett.119.157201,PhysRevX.8.031028} with the mixing of Mg$^{2+}$/Ga$^{3+}$~\cite{PhysRevLett.118.107202,paddison2016continuous}, as well as in other relevant materials~\cite{Kimchi2018Scaling}. Despite the growing interest in theory, the key issue is whether the paramagnetic phase conspired by frustration and randomness in real materials is relevant to the exotic QSL/RVB state with strong quantum fluctuations, or simply a trivial product state of quenched random singlets. To address this issue, local and dynamic measurements on QSL candidates with quantifiable randomness are particularly needed.

\begin{figure*}[t]
\begin{center}
\includegraphics[width=14cm,angle=0]{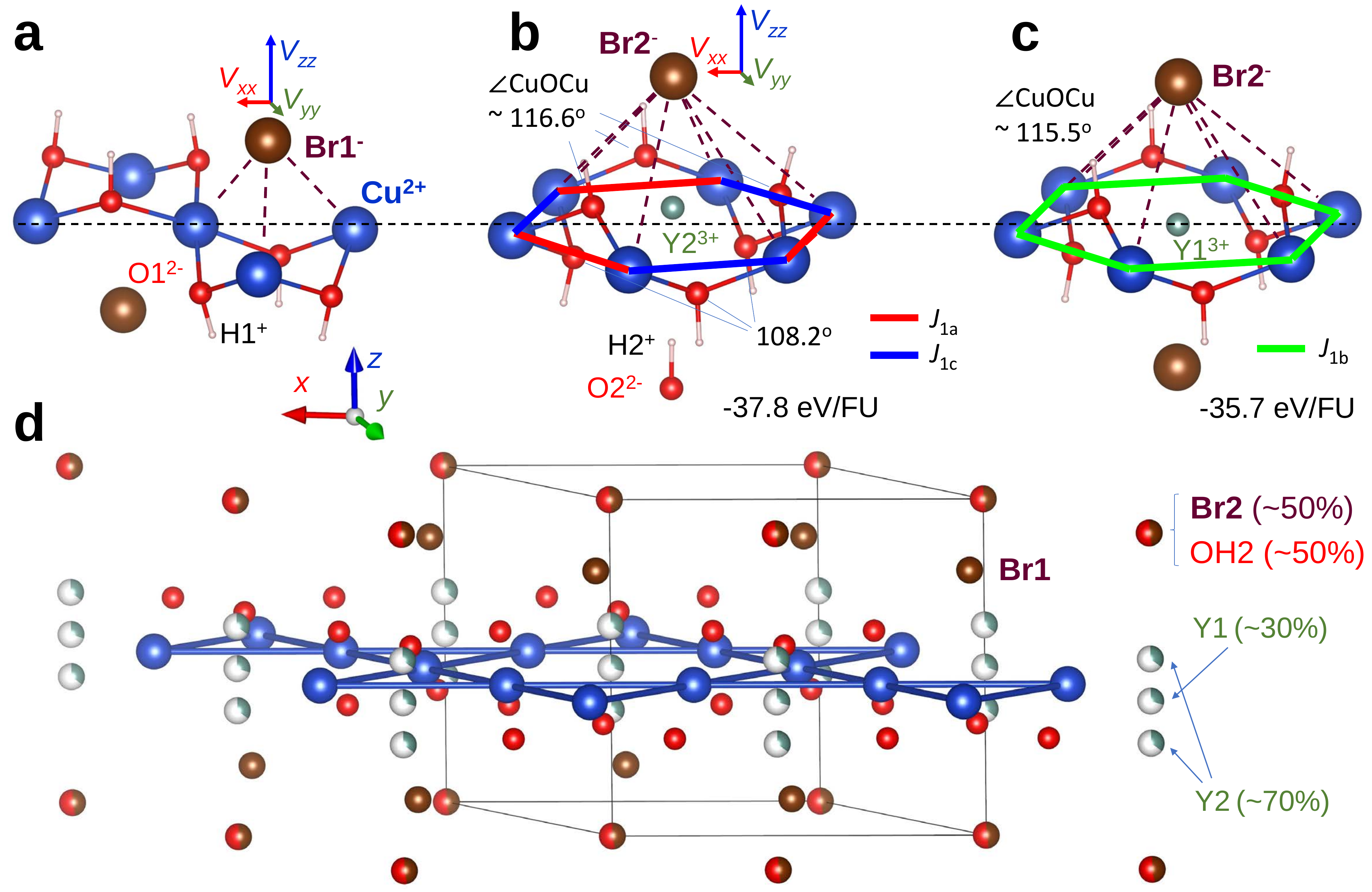}
\caption{\textbf{Crystal structure of YCu$_3$(OH)$_{6.5}$Br$_{2.5}$ around the kagome layer of Cu$^{2+}$.}
The Br1 nuclear ($^{81}$Br or $^{79}$Br ) spin detects three equidistant Cu$^{2+}$ electronic spins of each triangle on the kagome lattice (\textbf{a}), and the inset defines the coordinate system for the spin components.  Whereas, the Br2 nuclear spin mainly probes the nonsymmetric hexagon of spins with alternate exchanges (\textbf{b}), instead of the symmetric hexagon with almost uniform exchange (\textbf{c}). The formation energies of \textbf{b} and \textbf{c} stacking sequences are listed, and the different exchange paths of Cu-O-Cu ($J_{\textrm{1a}}$, $J_{\textrm{1b}}$, and  $J_{\textrm{1c}}$, depending on the bond angles) are marked. The principal axes of the electric field gradient calculated by density functional theory at the Br1 (\textbf{a}) and Br2 (\textbf{b}) sites are displayed by arrows scaled by the modulus of the tensor element,  $V_{zz}$ $\sim$ $-2V_{xx}$ $\sim$  $-2V_{yy}$  (see Supplementary Note 6).  In our measurements, the external magnetic field is always applied along the $z$ axis, and thus the second-order quadrupole shifts of the main NMR lines are negligibly small. (\textbf{d}) The crystal structure determined by the single-crystal x-ray diffraction~\cite{arXiv:2107.12712}. The thin lines mark the unit cell.}
\label{fig1}
\end{center}
\end{figure*}

Recently, a $S$ = 1/2 KHA YCu$_3$(OH)$_{6.5}$Br$_{2.5}$ (YCOB) has been proposed, without any global symmetry reduction of the kagome lattice (space group $P\bar{3}m1$, see Fig.~\ref{fig1}d)~\cite{CHEN2020167066,arXiv:2107.11942,arXiv:2107.12712}. Neither long-range magnetic ordering nor spin-glass freezing was observed down to 50 mK $\sim$ 0.001$\langle J_1\rangle$, as evidenced by specific heat~\cite{arXiv:2107.11942}, thermal conductivity~\cite{thermalCon}, and ac susceptibility~\cite{CHEN2020167066,arXiv:2107.12712} measurements. The observed power-law $T$ dependence of low-$T$ specific heat suggests the appearance of gapless spin excitations~\cite{arXiv:2107.11942,arXiv:2107.12712}. Unlike other known QSL materials (e.g. ZnCu$_3$(OH)$_6$Cl$_2$~\cite{PhysRevLett.100.087202,Fu2015Evidence,Khuntia2020Gapless,PhysRevLett.100.077203,PhysRevB.84.020411}), the mixing between Cu$^{2+}$ and other nonmagnetic ions is prohibited due to the significant ionic difference, thus defect orphan spins are essentially negligible~\cite{arXiv:2107.12712}. Further, the antisite mixing of the polar OH$^-$ and nonpolar Br$^-$ causes 70(2)\% of randomly distributed hexagons of alternate exchanges (e.g. Fig.~\ref{fig1}b) on the kagome lattice (Fig.~\ref{fig1}d),  which accounts for the measured thermodynamic properties above  $T\sim$ 0.1$\langle J_1\rangle$~\cite{arXiv:2107.12712}. Therefore, YCOB provides an excellent platform for a quantitative study of the aforementioned randomness-induced QSL physics.

Herein, the effect of inherent randomness on the QSL properties was investigated locally by $^{81}$Br and $^{79}$Br nuclear magnetic resonance (NMR) measurements on high-quality YCOB single crystals. We successfully identify the two NMR signals originating from the intrinsic kagome spins (Br1, see Fig.~\ref{fig1}a) and spins of hexagons with alternate exchanges on the kagome lattice (Br2, see Fig.~\ref{fig1}b). Simulations of the random exchange model show good agreement with the measured $T$ dependence of NMR line shifts and broadening, which suggests the formation of the randomness-induced QSL phase at low $T$. The measured in-plane spin fluctuations of the kagome spin system exhibit a weak power-law $T$ dependence down to $T$ $\sim$ 0.03$\langle J_1\rangle$ despite the quenched exchange randomness, thus supporting the survival of strong quantum fluctuations in YCOB.

\begin{figure*}[t]
\begin{center}
\includegraphics[width=13cm,angle=0]{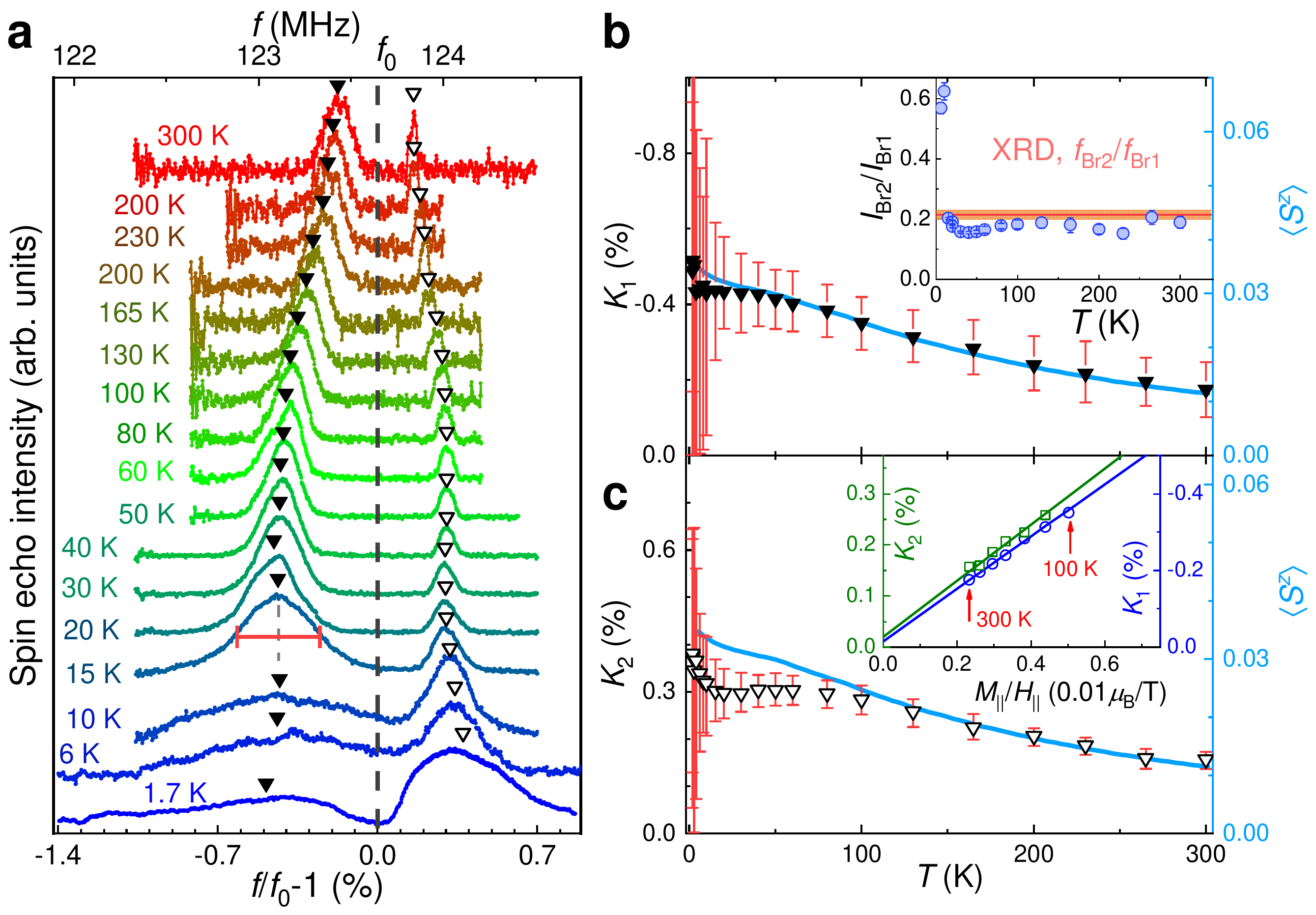}
\caption{\textbf{Nuclear magnetic resonance spectra of YCu$_3$(OH)$_{6.5}$Br$_{2.5}$.}
(\textbf{a}) Frequency-sweep spectra measured on the sample $S_1$ at a field $\mu_0H_{\parallel}\sim$ 10.75 T (the reference frequency $f_0$ = $^{81}\gamma_\mathrm{n}\mu_0H_{\parallel}\sim$ 123.64 MHz). The shifts of two lines, $K_1$ and $K_2$, are marked by solid and hollow triangles, respectively. Temperature dependence of $K_1$ (\textbf{b}) and $K_2$ (\textbf{c}), with the bulk magnetization ($\langle S^z\rangle$) measured at $\mu_0H_{\parallel}$ = 10.75 T for comparison. The inset of \textbf{b} shows the ratio between the integrated intensities of $^{81}$Br2 and $^{81}$Br1 lines ($I_{\mathrm{Br2}}/I_{\mathrm{Br1}}$), as well as the stoichiometric ratio from x-ray diffraction~\cite{arXiv:2107.12712}. Inset of \textbf{c} displays $K_1$ and $K_2$ shifts vs bulk susceptibility $\chi_{\parallel}$ (i.e. $M_{\parallel}/H_{\parallel}$). The red bars display the normalized frequency regions where the intensity is larger than half of the maximum value in \textbf{b} and \textbf{c}, and error bars on $I_{\mathrm{Br2}}/I_{\mathrm{Br1}}$  show a standard error from the fit.}
\label{fig2}
\end{center}
\end{figure*}

\bigskip \noindent {\bf Results}\\
\textbf{NMR spectra.}
Figure~\ref{fig2}a shows the $^{81}$Br NMR spectra measured on the crystal $S_1$ (Supplementary Fig. 1)  at $\mu_0H_{\parallel}\sim$ 10.75 T ($\sim$ 0.14$\langle J_1\rangle$). Two well-separated peaks are observed, originating respectively from two different Wyckoff positions of 2$d$ (Br1) and 1$a$ (Br2). Above $\sim$ 15 K, the ratio between the integrated intensities of these two peaks $I_{\mathrm{Br2}}/I_{\mathrm{Br1}}$ $\sim$ 0.2 is well consistent with the stoichiometric ratio of Br2 and Br1 determined by single-crystal x-ray diffraction (XRD) $f_{\mathrm{Br2}}/f_{\mathrm{Br1}}$ = 0.22(1)~\cite{arXiv:2107.12712} (see inset of Fig.~\ref{fig2}b). Below 15 K, the weight of Br1 NMR line decreases drastically and $I_{\mathrm{Br2}}/I_{\mathrm{Br1}}$ increases sharply, due to the reduced spin-spin relaxation times (Supplementary Note 4 and Supplementary Fig. 5)~\cite{PhysRevLett.116.107203}. Neither Br1 nor Br2 peak splits down to 1.7 K, suggesting the absence of conventional magnetic ordering within the ability of our resolution.

\begin{figure*}[t]
\begin{center}
\includegraphics[width=13cm,angle=0]{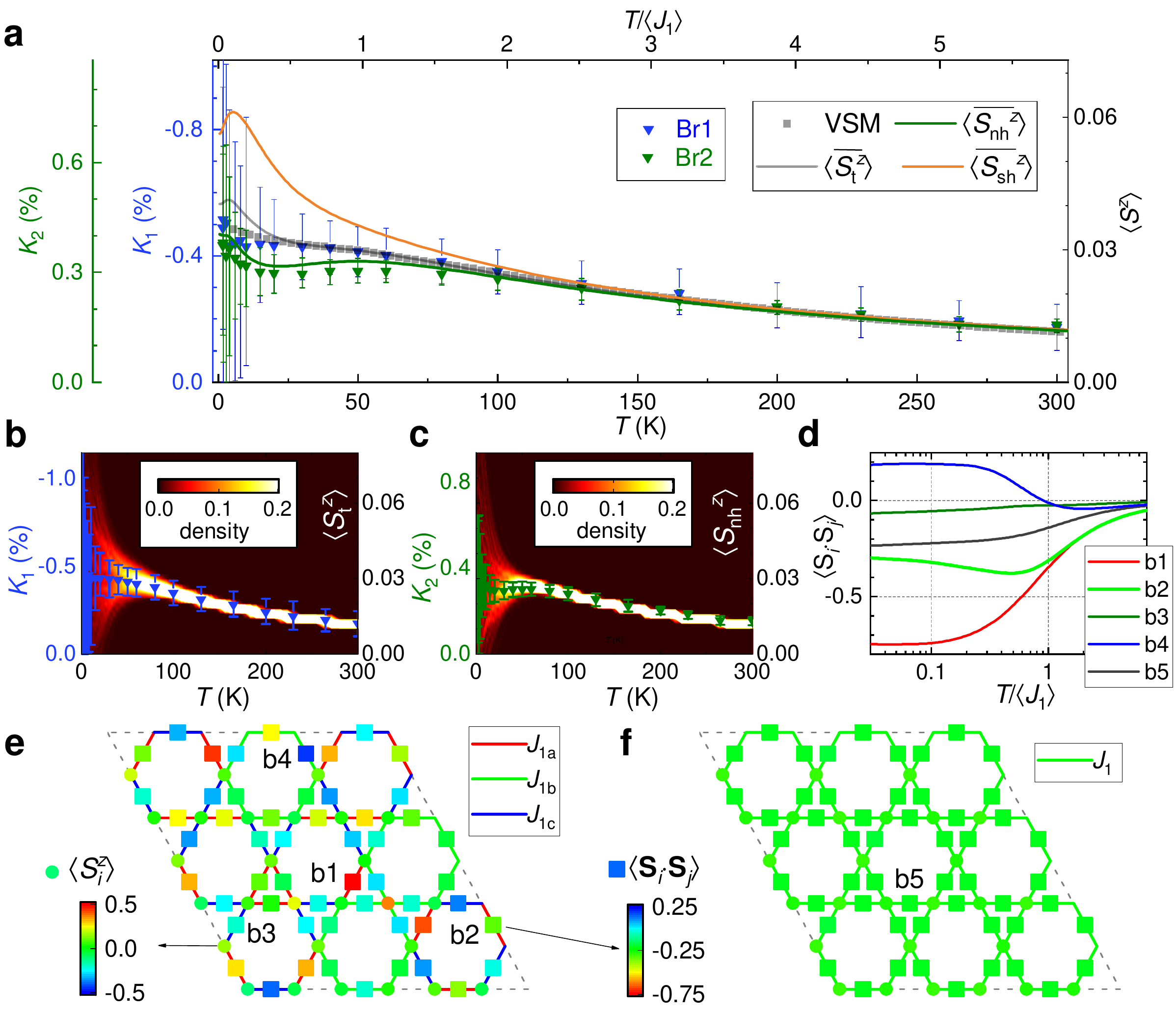}
\caption{\textbf{Simulations of the spatially inhomogeneous susceptibility probed by nuclear magnetic resonance (NMR) spectra.}
(\textbf{a}) Temperature dependence of NMR shifts $K_1$ and $K_2$, as well as bulk $\langle S^z\rangle$ measured in a vibrating sample magnetometer at a field $\mu_0H_{\parallel}$ = 10.75 T. The colored lines show the calculated average $\langle S^z\rangle$ over all the triangles ($\overline{\langle S_{\mathrm{t}}^z\rangle}$), nonsymmetric hexagons ($\overline{\langle S_{\mathrm{nh}}^z\rangle}$), and  symmetric hexagons ($\overline{\langle S_{\mathrm{sh}}^z\rangle}$). (\textbf{b},\textbf{c}) The finite-temperature Lanczos diagonalization results of local magnetization of triangles ($\langle S_{\textrm{t}}^z\rangle$) and nonsymmetric hexagons ($\langle S_{\textrm{nh}}^z\rangle$), along with the measured $K_1$ and $K_2$, respectively. (\textbf{d}) Calculated local correlations of the selected spin pairs (see \textbf{e},\textbf{f}) as function of normalized temperature $T/\langle J_1\rangle$. (\textbf{e}) Calculations of a sample within the random kagome Heisenberg antiferromagnet (KHA) model at $T$ = 0.1$\langle J_1\rangle$ and $\mu_0H_{\parallel}$ = 10.75 T. The solid circles and squares stand for local magnetization $\langle S_i^z\rangle$ at the kagome site $i$ and correlation function $\langle$\textbf{S}$_i$$\cdot$\textbf{S}$_j\rangle$ of the nearest-neighbor spin pair $\langle ij\rangle$, respectively. (\textbf{f}) The same calculations as in \textbf{e}, but within the ideal KHA model. The dashed lines mark the clusters with periodic boundary conditions, and $J_{\textrm{1a}}$, $J_{\textrm{1b}}$, $J_{\textrm{1c}}$, and $J_1$  present the exchange couplings. The color scale in \textbf{b} and \textbf{c} quantifies the distributed density, whereas the ones in \textbf{e} and \textbf{f} quantify the  local magnetization (circles) and correlation function (squares), respectively. The bars in \textbf{a}-\textbf{c} show the regions where the intensities are larger than half of the maximum value.}
\label{fig3}
\end{center}
\end{figure*}

The NMR shift of Br1 ($K_1$) detecting three ($z_{1}$ = 3) equidistant spins of each triangle on the kagome lattice (Fig.~\ref{fig1}a), follows the bulk susceptibility $\chi_{\parallel}$ measured at the same magnetic field strength in the full temperature range (Fig.~\ref{fig2}b), compatible with the absence of defect orphan spins in YCOB. In contrast, the NMR shift of Br2 ($K_2$) shows an obvious deviation from the bulk susceptibility below $\sim$ 100 K $\sim$ 2$\langle J_1\rangle$. Generically, the NMR shift consists of a $T$-dependent term proportional to the local susceptibility and a $T$-independent term ($K_{\mathrm{0}}$). Above 100 K, the NMR line broadening is insignificant (see Fig.~\ref{fig2}) and the calculated local magnetization is nearly spatially homogeneous (see Fig.~\ref{fig3}a), and thus one expects a scaling law $K$ = $A_{\mathrm{hf}}\chi_{\parallel}$+$K_{\mathrm{0}}$, where $A_{\mathrm{hf}}$ presents the hyperfine coupling between Br (Br1 or Br2) nuclear and Cu$^{2+}$ electronic spins. By fitting the experimental data (see inset of Fig.~\ref{fig2}c), we obtain $A_{\mathrm{hf1}}$ = $-$0.68(2) T/$\mu_\mathrm{B}$, $K_{\mathrm{01}}$ = $-$0.015(7)\%  and  $A_{\mathrm{hf2}}$ = 0.55(3) T/$\mu_\mathrm{B}$, $K_{\mathrm{02}}$ = 0.02(1)\%. The presence of both negative and positive hyperfine couplings of the same nuclear species is surprising, and the underlying mechanism must be complex, including both the positive and negative contributions.

The NMR shift of Br2 probes spins of hexagons on the kagome lattice (Fig.~\ref{fig1}b,c), but is obviously smaller than $A_{\mathrm{hf2}}\chi_{\parallel}$ ($K_{\mathrm{02}}$ $\ll$ $K_2$) below 100 K (Fig.~\ref{fig2}c). In fact, the formation energy of the optimized nonsymmetric Br2-OH2 stacking sequence (Fig.~\ref{fig1}b) ($-$37.8 eV/FU), is $\sim$ 2.5$\times$10$^4$ K/FU lower than that of the symmetric Br2-Br2 configuration (Fig.~\ref{fig1}c) ($-$35.7 eV/FU). Therefore, Br2 ions actually prefer the nonsymmetric local environments, and Br2 nuclear spins mainly probe the nonsymmetric hexagons with alternate exchanges~\cite{arXiv:2107.12712} ($J_{\mathrm{1a}}$ $>$ $J_{\mathrm{1c}}$) as illustrated in Fig.~\ref{fig1}b. Intuitively, the nonsymmetric hexagon tends to locally release the frustration and form three nonmagnetic singlets along the stronger couplings $J_{\mathrm{1a}}$ (see Fig.~\ref{fig3}e), which accounts for the relatively smaller Br2 shifts observed below 100 K (Fig.~\ref{fig3}a). Moreover, the Br1 line detects the site susceptibility/magnetization of all the Cu$^{2+}$ spins with exchange couplings $J_{\textrm{1a}}$, $J_{\textrm{1b}}$, and $J_{\textrm{1c}}$, whereas the Br2 line mainly probes the hexagons of spins only with $J_{\textrm{1a}}$ and $J_{\textrm{1c}}$. The Br2 line probes less kinds of Cu$^{2+}$ spins, and thus is narrower than the Br1 line.

Quantitatively, the temperature dependence of both Br1 and Br2 shifts can be reproduced by the average magnetization of all the triangles and nonsymmetric hexagons on the kagome lattice, viz. $K_1$ $\sim$ $A_{\mathrm{hf1}}g_{\parallel}\overline{\langle S_{\textrm{t}}^z\rangle}/(\mu_0H_{\parallel})$ and  $K_2$ $\sim$ $A_{\mathrm{hf2}}g_{\parallel}\overline{\langle S_{\textrm{nh}}^z\rangle}/(\mu_0H_{\parallel})$, respectively (see Fig.~\ref{fig3}a), with $J_{\textrm{1a}}$ = 89 K, $J_{\textrm{1b}}$ = 48 K, and  $J_{\textrm{1c}}$ = 16 K ($\langle J_1\rangle\sim$ 51 K) experimentally determined by bulk susceptibilities (Supplementary Note 1 and Supplementary Fig. 2). Taking all the triangles and nonsymmetric hexagons into account, we are able to simulate the broadening of both Br1 and Br2 lines by introducing the distributed density $\propto$ d$n_S(\langle S^z\rangle)$/d$\langle S^z\rangle$ (see Fig.~\ref{fig3}b,c), where d$n_S(\langle S^z\rangle)$ is the number of triangles or nonsymmetric hexagons with the local magnetization (per site) ranging from $\langle S^z\rangle$ to $\langle S^z\rangle$+d$\langle S^z\rangle$. $\langle S^z\rangle$ is thermally averaged, so the distributed density is a function of $T$.

The simulations also enable us to revisit the correlation functions $\langle$\textbf{S}$_i$$\cdot$\textbf{S}$_j\rangle$ in this KHA system with randomness. Compared to the ideal case (Fig.~\ref{fig3}f), a small fraction ($\sim$ 3/54) of well-defined singlets with $\langle$\textbf{S}$_i$$\cdot$\textbf{S}$_j\rangle$ $\sim$ $-$0.7 $\to$ $-$0.75 are frozen at low $T$ in YCOB (Fig.~\ref{fig3}e), which is a signature of releasing frustration due to the quenched randomness.  These local singlets might confine the mobile spinons~\cite{PhysRevLett.104.177203,PhysRevLett.122.137201,Li2019YbMgGaO4,Spin2020li}, which might be responsible for the absence of large magnetic thermal conductivity observed in nearly all of the existing gapless QSL candidates, including the well-known ZnCu$_3$(OH)$_6$Cl$_2$~\cite{huang2021heat,murayama2021universal}, EtMe$_3$Sb[Pd(dmit)$_2$]$_2$~\cite{PhysRevLett.123.247204,PhysRevX.9.041051}, etc. However, the majority of antiferromagnetic interactions remain not fully satisfied at low $T$ (see Fig.~\ref{fig3}d,e), and the GS wavefunction should be represented by a superposition of various pairings of spins. It is worth to mention that the weights of different pairings should be different due to the quenched randomness.  The survival of strong frustration in the $S$ = 1/2 random KHA speaks against the product GS wavefunction of randomly distributed singlets, and may still give rise to strong quantum fluctuations. To testify this, we turn to the spin dynamics of YCOB mainly probed by the spin-lattice relaxation rates as follow.

\begin{figure*}[t]
\begin{center}
\includegraphics[width=17cm,angle=0]{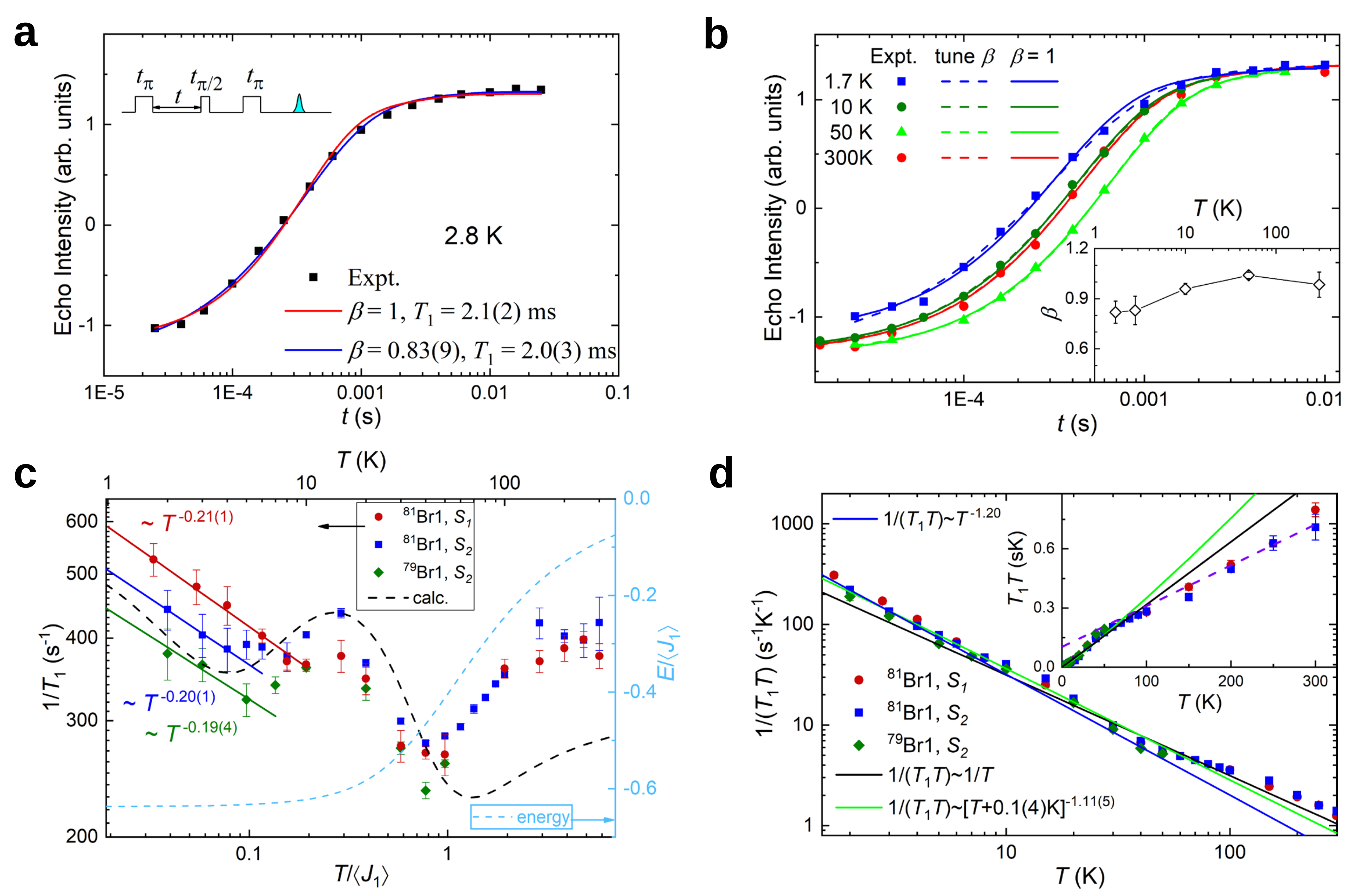}
\caption{\textbf{Nuclear spin-lattice relaxation of YCu$_3$(OH)$_{6.5}$Br$_{2.5}$.}
(\textbf{a}) A representative $T$ = 2.8 K spin-lattice relaxation of Br1 measured by an inversion recovery method, fitted to the stretched-exponential function with fixing $\beta$ = 1 (red) and tuning  $\beta$ (blue). The pulse sequence for $T_1$ measurements is depicted in the inset. (\textbf{b}) The same fits to the relaxation data measured at other selected temperatures. The inset shows the fitted stretching exponent $\beta$. (\textbf{c}) Temperature dependence of $^{81}$Br1 nuclear spin-lattice relaxation rate $1/T_1$ measured on the sample $S_1$ at a field $\mu_0H_{\parallel}$ = 10.75 T, as well as $^{81}$Br1 and $^{79}$Br1 $1/T_1$ measured on $S_2$ at $\mu_0H_{\parallel}$ = 10.75 and 11.59 T, respectively. The colored lines present the power-law fits to the experimental data below $\sim$ 10 K $\sim$ 0.2$\langle J_1\rangle$, the dashed black and blue lines are the $1/T_1$ and energy (per site), respectively, calculated by using the random kagome Heisenberg antiferromagnet model of YCu$_3$(OH)$_{6.5}$Br$_{2.5}$. (\textbf{d}) The Curie ($\sim$ $T^{-1}$, black lines) and critical ($\sim$ ($T-T_\mathrm{c}$)$^{-\alpha'-1}$, the critical temperature $T_\mathrm{c}$ = $-$0.1$\pm$0.4 K, green lines) fits to $1/(T_1T)$ at 1.7 $\leq$ $T$ $\leq$ 300 K. The blue line is the low-$T$ power-law dependence as shown in \textbf{c}. The inset shows the $T_1T$ vs $T$ plot, where the dashed violet line displays the antiferromagnetic Curie-Weiss behavior ($T_1T$ $\sim$ $T$+$\langle J_1\rangle$, with coupling $\langle J_1\rangle$ $\sim$ 50 K). The $T_1$ data presented in \textbf{c} and \textbf{d} are obtained from the single-exponential fits (i.e., $\beta$ = 1, see \textbf{a} and \textbf{b}), and the stretched-exponential fits are made only for comparison in \textbf{a} and \textbf{b}. Error bars on the experimental data points show a standard error from fit, and the error bars in \textbf{a} and \textbf{b} are small.}
\label{fig4}
\end{center}
\end{figure*}

\bigskip \noindent {\bf Spin dynamics.} The representative spin-lattice relaxation data measured on YCOB are displayed in Figs~\ref{fig4}a and \ref{fig4}b, which can be well fitted to the single-exponential function for the central transition of $I$ = 3/2 nuclear spins, i.e.,
\begin{equation}
M(t)=M_0 -2M_0 F (\frac{1}{10}\mathrm{e}^{-\frac{t}{T_1}}+\frac{9}{10}\mathrm{e}^{-\frac{6t}{T_1}}),
\label{Eqn1}
\end{equation}
where $M_0$ and $F$ are scale parameters for intensity. Alternatively, the relaxation data may be fitted with the stretched-exponential one,
\begin{equation}
M(t)=M_0 -2M_0 F [\frac{1}{10}\mathrm{e}^{-(\frac{t}{T_1})^\beta}+\frac{9}{10}\mathrm{e}^{-(\frac{6t}{T_1})^\beta}].
\label{Eqn2}
\end{equation}
Here, the stretching exponent $\beta$ slightly decreases at low temperatures (see inset of Fig.~\ref{fig4}b), but remains large down to the lowest temperature of 1.7 K, $\beta$ $>$ 0.8. Moreover, the fit to the single $T_1$ function (i.e. $\beta$ = 1) is still good even at 1.7 K (Fig.~\ref{fig4}b), with the adj. $R^2$ = 0.997~\cite{web}. The inclusion of the additional fitting parameter $\beta$ only slightly improves the fit (the adj. $R^2$ increases to 0.998), and even makes the standard error on $T_1$ larger. Therefore, all the following $T_1$ data  are obtained with the single-exponential fits.

Figure~\ref{fig4}c shows the results of Br1 nuclear spin-lattice relaxation rates (1/$T_1$). Br1  1/$T_1$ is highly sensitive to the electronic spin fluctuations perpendicular to the applied magnetic field on the kagome layer via the hyperfine coupling~\cite{2014Quantumspin,PhysRevLett.109.227004,PhysRevB.103.014431,PhysRevX.10.011007},
\begin{multline}
\frac{1}{T_1}\sim\frac{T}{hf_0}\sum_{\bm{q}}\chi''(\bm{q},f_0)\sim\frac{\pi\gamma_\mathrm{n}^2A_{\mathrm{hf1}}^2g_{\perp}^2}{Z}\sum_{\bm{q},m,m'}\mathrm{e}^{-\frac{E_m}{k_{\mathrm{B}}T}}\\
\times|\langle m'| S_{\bm{q}}^{\perp}|m\rangle|^2\delta (f_0-\frac{E_{m'}-E_m}{h}),
\label{e1}
\end{multline}
where the $\bm{q}$ dependence of $A_{\mathrm{hf1}}$ is  neglected, $f_0$ = $\gamma_\mathrm{n}\mu_0H_{\parallel}$ ($\ll$ $k_{\mathrm{B}}T/h$) $\to$ 0 presents the NMR frequency, $S_{\bm{q}}^{\perp}$ is the Fourier transform of $S_{i}^{\perp}$ over all triangles, and $Z$ = $\sum_{m}\exp (-E_m/k_{\mathrm{B}}/T)$ the partition function. In our simulation of $1/T_1$, the delta function is replaced by a Gaussian distribution with narrow width $\sim$ 10$^{-4}\langle J_1\rangle/h$~\cite{2014Quantumspin,PhysRevX.10.011007}.

At high temperatures ($T$ $\gg$ $\langle J_1\rangle$), the Moriya paramagnetic limit yields the $T$-independent $1/T_1^\infty$ = $(2\pi)^{1.5}\gamma_\mathrm{n}^2A_{\mathrm{hf1}}^2g_{\perp}^2S(S+1)/(3z_{1}\nu_e)$ $\sim$ 300 s$^{-1}$, reasonably comparable with the observation (see Fig.~\ref{fig4}c), where $\nu_e$ = $\langle J_1\rangle\sqrt{2zS(S+1)/3}/h$ is the exchange frequency with the coordination number $z$ = 4~\cite{Moriya1956Nuclear,PhysRevLett.127.157202,PhysRevB.103.014431}. Upon cooling, 1/$T_1$ slightly decreases first and then rises at $T$ $\sim$ $\langle J_1\rangle$. These features are qualitatively reproduced by the random KHA model of YCOB (Fig.~\ref{fig4}c). It is challenging to precisely simulate $1/T_1$, possibly due to the neglecting  of the $\bm{q}$ dependence of $A_{\mathrm{hf}}$~\cite{PhysRevB.103.014431}. At $T$ = 15 K $\sim$ 0.3$\langle J_1\rangle$, a weak anomaly (kink) of $1/T_1$ is observed, coinciding with the saturation of local nearest-neighbor correlations $\langle$\textbf{S}$_i$$\cdot$\textbf{S}$_j\rangle$ (see Fig.~\ref{fig3}d). This is possibly attributed to the emergence of short-range spin correlations, as seen in $\kappa$-(ET)$_2$Cu$_2$(CN)$_3$, PbCuTe$_2$O$_6$, etc.~\cite{PhysRevLett.115.077001,PhysRevB.73.140407,PhysRevLett.116.107203,PhysRevB.103.214405}.

\bigskip \noindent {\bf Discussion}\\
As $T$ further decreases below $\sim$ 10 K $\sim$ 0.2$\langle J_1\rangle$, $1/T_1$ exhibits a weak enhancement, $1/T_1\sim T^{-\alpha}$ with $\alpha$ $\sim$ 0.20$\pm$0.02, at least down to 1.7 K $\sim$ 0.03$\langle J_1\rangle$. This weak enhancement of $1/T_1$ is well reproducible between different single-crystal samples, nuclear spin probes ($^{81}$Br and $^{79}$Br), and applied magnetic fields ($\mu_0H_{\parallel}$ = 10.75 and 11.59 T), as shown in Fig.~\ref{fig4}. It is necessary to mention that the similar slowing down of spin fluctuations was also reported in other quantum disordered spin systems~\cite{PhysRevX.4.031008,lee2020emergent}. 
Such a behavior of $1/T_1$ suggests the spin system of YCOB may be proximate to a gapless/critical QSL state or highly dynamic valence bond glass with long-range fluctuating singlets:

First, such a temperature dependence of $1/T_1$ is inconsistent with conventional glassy spin freezing. Glassy freezing is typically observed by NMR as a broad peak in 1/$T_1$ vs $T$ when the inverse correlation time matches the NMR frequency, and such a peak ($T_1$ minimum) defines the freezing temperature $T_\mathrm{c}$~\cite{PhysRevLett.83.604,PhysRevB.76.054452,frachet2020hidden}. Moreover, 1/$T_1$ is usually expected to increase by more than one order of magnitude at $T_\mathrm{c}$ from that above $T_\mathrm{c}$~\cite{PhysRevLett.83.604,PhysRevB.76.054452,frachet2020hidden}. However, such a robust peak in 1/$T_1$ vs $T$ is absent in YCOB, as shown in Fig.~\ref{fig4}c. Other NMR quantities, line width and 1/$T_2$ (Supplementary Fig. 6), also exhibit the common features of well-reported QSL candidates~\cite{PhysRevB.77.104413,PhysRevB.103.024413,PhysRevLett.116.107203,Fu2015Evidence,PhysRevB.73.140407,PhysRevB.93.214432}, and speak against the existence of well-defined $T_\mathrm{c}$ in YCOB (Supplementary Note 5). Note that other experiments, including magnetic and thermodynamic measurements~\cite{CHEN2020167066,arXiv:2107.11942,arXiv:2107.12712}, also have precluded the possibility of conventional magnetic transition. Thereby, the formation of short-range spin correlations  (i.e., $\langle$\textbf{S}$_i$$\cdot$\textbf{S}$_j\rangle$ $<$ 0) should be responsible for the broadening of NMR lines and enhancement of 1/$T_1$ and 1/$T_2$ observed at low temperatures in YCOB.

Second, classical spin fluctuations are driven by thermal energy $k_\mathrm{B}T$, and cease at low temperatures, $T$ $\ll$ $\langle J_1\rangle$~\cite{balents2010spin}. A real example can be the $S$ = 5/2 KHA Li$_9$Fe$_3$(P$_2$O$_7$)$_3$(PO$_4$)$_2$, where the slowing down of classical spin fluctuations, 1/$T_1$ $\sim$ $T^{-1}$, observed just above the antiferromagnetic transition temperature~\cite{PhysRevLett.127.157202}, increases much faster than that seen in YCOB, upon cooling. Therefore, the observation of 1/$T_1$ $\sim$ $T^{-0.20\pm0.02}$ suggests that quantum fluctuations must play an important role at low temperatures ($T$ $\le$ 0.1$\langle J_1\rangle$) in YCOB.

Third, below $T$ $\sim$ 0.1$\langle J_1\rangle$, the calculated local nearest-neighbor correlations (Fig.~\ref{fig3}d) for the spin system of YCOB level off, manifesting that the observed quantum critical slowing down of spin fluctuations (i.e., $1/T_1\sim (T-T_\mathrm{c})^{-0.20\pm0.02}$ with $T_\mathrm{c}$ $\sim$ 0 K) are associated with a gapless nature of the spin excitations. This is further supported by the nearly quadratic $T$ dependence of specific heat ($C\sim T^{2.31}$)~\cite{arXiv:2107.12712}. In contrast, an exponential $T$-dependent decrease of $1/T_1$ and $C$ with a spin gap of $\geq$ $J_{\textrm{1c}}$ $\sim$ 16 K would be expected upon cooling, if the simple product GS of fully quenched random nearest-neighbor singlets is established at low $T$.

Finally, the system energy gets nearly $T$-independent (Fig.~\ref{fig4}c) and the measured residual spin entropy is small $\sim$ 3.4\%$R$ln2~\cite{arXiv:2107.12712}, indicating approaching to the disordered GS properties of YCOB at $\sim$ 1.7 K. We notice that although the temperature decreases by more than two orders of magnitude from 300 to 1.7 K, 1/$T_1$ only slightly increases from $\sim$ 380 to 500 s$^{-1}$, thus suggesting the survival of $\sim$ 70\% of total spin fluctuations at low temperatures (i.e. quantum fluctuations)~\cite{PhysRevLett.109.227004,PhysRevLett.117.097201,PhysRevB.103.024413,lee2020emergent}. Since $1/T_1$ only weakly depends on temperature in the full temperature range between 1.7 and 300 K, $1/(T_1T)$ $\propto$ $\sum_{\bm{q}}\chi''(\bm{q},f_0)$ (see equation~(\ref{e1})) follows a Curie behavior diverging towards $T$ = 0 K, as shown in Fig.~\ref{fig4}d. We also fit the data with the critical function, 1/($T_1T$) $\sim$ ($T-T_\mathrm{c}$)$^{-\alpha'-1}$, in the full temperature range, and find $T_\mathrm{c}$ = $-$0.1$\pm$0.4 K $\sim$ 0 K and $\alpha'$ = 0.11(5) (see the green lines in Fig.~\ref{fig4}d). The dynamical susceptibility ($\sum_{\bm{q}}\chi''(\bm{q},f_0)$) relates to the second derivative of the free energy, and thus the observation of $T_\mathrm{c}$ $\sim$ 0 K (Fig.~\ref{fig4}d) indicates a quantum critical behavior. In the paramagnetic limit ($T$ $\to$ $\infty$), the spin fluctuations including the extremely strong thermal fluctuations ($\propto$ $T$, see ref.~\cite{balents2010spin}) take huge values for an arbitrary spin system. In YCOB, nearly the same Curie behavior, 1/$(T_1T)$ $\propto$ $\sum_{\bm{q}}\chi''(\bm{q},f_0)$ $\propto$ 1/$T$, persists from 300 K $\gg$ $\langle J_1\rangle$ down to 1.7 K $\sim$ 0.03$\langle J_1\rangle$ (see Fig.~\ref{fig4}d),  thus supporting the survival of strong spin fluctuations toward zero temperature (i.e. quantum fluctuations). All together, these features are in line with the spin dynamics of the putative gapless/critical QSL or dynamic valence bond glass with long-range fluctuating singlets.

We have investigated the local spin susceptibility and dynamics by Br NMR measurements on the high-quality single crystals of YCOB whose randomness has been quantified. The quenched exchange randomness gives rise to the spatially inhomogeneous susceptibility, which accounts for the different $T$ dependencies of Br1 and Br2 main line shifts, as well as broadening. Despite a small fraction of frozen random singlets, the majority of spins of YCOB evade conventional magnetic ordering and remain highly fluctuating, as evidenced by the weak power-law $T$ dependence of $1/T_1$ down to $T$ $\sim$ 0.03$\langle J_1\rangle$. Our work highlights the role of quantum fluctuations to the gapless QSL behaviors generally observed in relevant materials with inevitable randomness.

\bigskip \noindent {\bf Methods}\\
\textbf{NMR measurements.} Two single-crystal samples of YCOB ($S_1$ and $S_2$ with $\sim$ 6.2 and 15.2 mg, respectively) used in NMR measurements were grown by a recrystallization in a temperature gradient~\cite{arXiv:2107.12712}. The $^{81}$Br ($^{81}\gamma_\mathrm{n}$  = 11.4989 MHz/T) and $^{79}$Br ($^{79}\gamma_\mathrm{n}$  = 10.6675 MHz/T) NMR measurements were performed using standard spin echo sequences at external magnetic fields $\sim$10.75 and 11.59 T aligned to the $c$ axis, between 300 and 1.7 K (the base temperature of our setup). Main NMR shifts of both Br1 and Br2 are derived from the central transitions recorded by a stepped frequency sweep ranging 121-125 MHz wherein no satellite transitions from Br can be observed due to the large nuclear quadrupole frequency $\nu_{Q}$ $>$ 20 MHz (Supplementary Note 2 and Supplementary Fig. 3). However, the element of the electric field gradient (EFG) tensor $V_{zz}$ is nearly parallel to the external field, and thus the second-order quadrupole shift is negligibly small (Supplementary Note 6 and Supplementary Figs 7 and 8). Spin-lattice ($1/T_1$) and spin-spin ($1/T_2$) relaxation rates are investigated on Br1 site, measured in inversion recovery and Hahn spin-echo decay methods, respectively (Supplementary Note 3 and Supplementary Fig. 4). The error bars on $T_1$ come from the standard nonlinear curve fits by using the Origin program, as shown in Fig.~\ref{fig4}.

\bigskip \noindent {\bf Simulations.} We conducted the finite-temperature Lanczos diagonalization simulations of the random KHA model~\cite{arXiv:2107.12712} for the NMR line shifts and broadening, local magnetization and correlation, as well as $1/T_1$. No significant finite-size effect of the calculation was observed down to $T$ $\sim$ 0.1$\langle J_1\rangle$~\cite{arXiv:2107.12712}. The formation energy (see Fig.~\ref{fig1}) is calculated as $E$(YCu$_3$(OH)$_{6+x}$Br$_{3-x}$) $-\mu$(Y)$-3\mu$(Cu)$-$(6$+x$)$\mu$(OH)$-$(3$-x$)$\mu$(Br), where the total energy is obtained from the previous density functional theory calculation~\cite{arXiv:2107.12712}, and $\mu$(Y), $\mu$(Cu), $\mu$(OH), $\mu$(Br) are the chemical potentials of the constituents. The international system of units is used throughout this paper, and $\langle \rangle$ presents thermal and sample average.

\bigskip \noindent {\bf Data availability.} \\
The data sets generated during and/or analysed during the current study are available from the corresponding author on reasonable request.

\bigskip \noindent {\bf References} \\

\bigskip \noindent {\bf Acknowledgements}\\
We thank Philipp Gegenwart, Alexander A. Tsirlin,  Christian Hess, Xiaochen Hong, and Haijun Liao  for insightful discussions, and Long Ma for technical helps. This work was supported by the Fundamental Research Funds for the Central Universities (HUST: 2020kfyXJJS054) and the open research fund of Songshan Lake Materials Laboratory (2022SLABFN27).

\bigskip \noindent {\bf Author contributions}\\
Y.K.L. and Y.S.L. planned the experiments. L.Y., B.Q.L. and Y.S.L. synthesized and characterized the sample. F.J.L., J.Z. and Y.K.L. collected the NMR data. F.J.L., Y.K.L. and Y.S.L. analysed the data. Y.S.L. conducted the quantum many-body and first-principles computations. Y.S.L. and Y.K.L. wrote the manuscript with comments from all co-authors. The manuscript reflects the contributions of all authors.


\bigskip \noindent {\bf Competing interests} \\
The authors declare no competing interests.

\clearpage

\addtolength{\oddsidemargin}{-0.75in}
\addtolength{\evensidemargin}{-0.75in}
\addtolength{\topmargin}{-0.725in}

\newcommand{\addpage}[1] {
 \begin{figure*}
   \includegraphics[width=8.5in,page=#1]{Supplementary_r2.pdf}
 \end{figure*}
}
\addpage{1}
\addpage{2}
\addpage{3}
\addpage{4}
\addpage{5}
\addpage{6}
\addpage{7}
\addpage{8}
\addpage{9}
\addpage{10}
\addpage{11}
\addpage{12}

\end{document}